\documentstyle[11pt,epsfig]{article}

\begin{document}
\large

\def\lsim{\mathrel{\rlap{\lower3pt\hbox{\hskip0pt$\sim$}}
    \raise1pt\hbox{$<$}}}         
\def\gsim{\mathrel{\rlap{\lower4pt\hbox{\hskip1pt$\sim$}}
    \raise1pt\hbox{$>$}}}         
\def\dblint{\mathop{\rlap{\hbox{$\displaystyle\!\int\!\!\!\!\!\int$}}
    \hbox{$\bigcirc$}}}
\def\ut#1{$\underline{\smash{\vphantom{y}\hbox{#1}}}$}

\newcommand{\beq}{\begin{equation}}
\newcommand{\eeq}{\end{equation}}
\newcommand{\dem}{\Delta M_{\mbox{B-M}}}
\newcommand{\dega}{\Delta \Gamma_{\mbox{B-M}}}

\newcommand{\ind}[1]{_{\begin{small}\mbox{#1}\end{small}}}
\newcommand{\WA}{{\em WA}}
\newcommand{\SM}{Standard Model }
\newcommand{\QCD}{{\em QCD }}
\newcommand{\KM}{{\em KM }}
\newcommand{\hscale}{\mu\ind{hadr}}
\newcommand{\sG}{i\sigma G}

\newcommand{\MS}{\overline{\mbox{MS}}}
\newcommand{\pole}{\mbox{pole}}
\newcommand{\aver}[1]{\langle #1\rangle}

\newcommand{\appa}{\mbox{\ae}}
\newcommand{\CP}{{\em CP } }
\newcommand{\fy}{\varphi}
\newcommand{\hi}{\chi}
\newcommand{\al}{\alpha}
\newcommand{\as}{\alpha_s}
\newcommand{\gf}{\gamma_5}
\newcommand{\gm}{\gamma_\mu}
\newcommand{\gn}{\gamma_\nu}
\newcommand{\be}{\beta}
\newcommand{\ga}{\gamma}
\newcommand{\de}{\delta}
\renewcommand{\Im}{\mbox{Im}\,}
\renewcommand{\Re}{\mbox{Re}\,}
\newcommand{\GeV}{\,\mbox{GeV}}
\newcommand{\MeV}{\,\mbox{MeV}}
\newcommand{\matel}[3]{\langle #1|#2|#3\rangle}
\newcommand{\state}[1]{|#1\rangle}
\newcommand{\ra}{\rightarrow}
\newcommand{\ve}[1]{\vec{\bf #1}}

\newcommand{\rhs}{{\em rhs}}
\newcommand{\pp}{\langle \ve{p}^2 \rangle}

\newcommand{\BR}{\,\mbox{BR}}
\newcommand{\La}{\overline{\Lambda}}

\vspace*{.7cm}
\begin{flushright}
\large{
CERN-TH.7282/94\\
UND-HEP-94-BIG\hspace*{0.1em}06}\\
\end{flushright}
\vspace{1.2cm}
\begin{center} \LARGE {\bf BEAUTY PHYSICS IN THE
NEXT 15 YEARS -- \\
AN ITINERARY FOR A LONG JOURNEY
TOWARDS AN ESSENTIAL GOAL
\footnote{Theory Review talk given at
{\bf Beauty '94}, Le Mont Saint Michel, France,
April 1994}}
\end{center}
\vspace*{.4cm}
\begin{center} \Large
I.I. Bigi\\
\vspace*{.4cm}
{\normalsize {\it Theoretical Physics Division, CERN, CH-1211
Geneva 23, Switzerland \footnote{During the academic year
1993/94}}\\
and\\
{\it Dept. of Physics,
University of Notre Dame du
Lac, Notre Dame, IN 46556 \footnote{Permanent address} } \\
{\it e-mail address: IBIGI@VXCERN, BIGI@UNDHEP.HEP.ND.EDU}}
\vspace{.4cm}
\end{center}
\thispagestyle{empty} \vspace{.4cm}

\centerline{\Large\bf Abstract}
\vspace{.4cm}
After re-stating why beauty physics is so special and
sketching its present status I attempt to map out an
itinerary for the next fifteen years. The basic elements
are listed for a research program that allows to define and
probe the KM unitarity triangle in a manner where the
numerical uncertainties are reduced to the few percent
level. Dedicated experiments at the LHC will be
essential in exploiting the discovery guarantee in
beauty physics to the fullest.

\vspace{3cm}
{\em This paper is dedicated to Prof. H. Kastrup on the
occasion of his 60th birthday. I have spent five years at
his institute at the RWTH Aachen
and I truly enjoyed the stimulating atmosphere he created by
encouraging intellectual exchange unimpeded by
hierarchical privileges.}

\vspace{2cm}
\noindent CERN-TH.7282/94 \\
\noindent May 1994
\newpage
\large
\addtocounter{footnote}{-3}
\addtocounter{page}{-1}

Approaching Mont Saint Michel one recognizes the church
crowning the rock from far away. Getting closer one discovers
there is much more structure to the Mont; after having spent
a little while here one realizes that these
other buildings do more than
merely support the church, that they hold their own charm
and beauty. At the end of my talk I will return to this image
and explain why it provides us
which such a fitting allegory on the
purpose of our meeting.

I will argue that a satisfactory
research program in beauty physics requires and deserves
a long term commitment. Before we set out on a long and
arduous journey it behooves us to recall past experiences
to prepare our souls and spirits for the adventures that lie
ahead. So let us remember the main lessons we have learnt
from the study of the production and decay of strange hadrons:
the full relevance of the concepts of flavour quantum numbers
and of Cabibbo universality was revealed; the first evidence for
parity violation emerged through the $\theta -\tau$ puzzle;
the tremendous suppression of flavour-changing neutral
currents became apparent leading to the prediction of a
then new quantum number, namely charm, including its
mass scale; the discovery of CP violation fundamentally
changed our perception of nature's microscopic design
and, on a more practical level,
showed the existence of New Physics even beyond a
two-family ansatz. All these insights were essential
ingredients in formulating what we call now the
Standard Model (SM) -- although it has to be said that
at first they were often met with considerable skepticism.

While this Standard Model has so far yielded a consistent
description, it is generally viewed as incomplete. This
is based on the realization that it contains various mysteries.
Central among them is the origin of mass in general and of
fermions in particular, the existence of families and the
very peculiar form of the KM matrix. CP violation is intimately
connected with these mysterious features.

My basic contention is: a detailed and comprehensive analysis
of beauty decays that includes meaningful studies of CP violation

\noindent $\bullet$ provides novel and unique perspectives
on fundamental mysteries of SM,

\noindent $\bullet$ will very likely reveal New Physics (NP) and

\noindent $\bullet$ will possibly be essential in formulating
the `New Standard Model'.

In the remainder of my talk I will state `why beauty decays' are
so special, summarize the `status quo', sketch the `next five
years', rhapsodize about the `heroic period' and point to the
`ultimate measurements' before concluding with an `outlook'.

\section{\bf Why Beauty?}

The lifetimes of beauty hadrons are of order 1 psec, i.e. they
are `long' according to two criteria:
(i) Decay vertices of beauty
hadrons can be resolved with available technologies once they
move with sufficient velocity.
(ii) Beauty decays are much more than
Cabibbo suppressed: $|V(cb)|^2\ll \sin ^2\theta _C^2$. Since
SM tree level transitions are thus reduced, the sensitivity to
quantum effects and/or NP is considerably enhanced.

SM quantum effects are actually GIM enhanced since
$m_{top} > M_W$. This generates `speedy'
$B^0-\bar B^0$ oscillations --
$\Delta m(B)/\Gamma (B) \geq {\cal O}(1)$ -- and a relatively
large branching ratio for $B\ra K^*\gamma $. As we have heard
from Ali at this meeting \cite{ALI}
there exists an extremely rich
phenomenology of B decays involving all three
families, namely in semi-leptonic, radiative and other
rare transitions.

The most fascinating aspect is that SM predicts, as reviewed
here by Wyler \cite{WYLER},
huge CP asymmetries of order few$\times$10\%
in some B decays due to the undiluted interplay of all
three families and the high $B^0-\bar B^0$ oscillation rate.

This exciting phenomenology is predicted with fairly high
theoretical reliability in some relevant cases at least.
As discussed by Berger \cite{BERGER},
production rates for beauty hadrons
should be predicted with reasonable accuracy. For some ratios
of transition rates one can expect the major part of the
hadronic uncertainties to drop out, like in
$\Delta m(B_s)/\Delta m(B_d)
\simeq |V(ts)|^2/|V(td)|^2$.
Other decay rates can be expressed in terms of the
basic parameters $V(cb)$, $V(ub)$ , $V(td)$ etc. with
at least decent accuracy. Furthermore one can translate these
parametric predictions into numerical ones by employing
$1/m_b$ expansions to extract the values of these parameters
from semileptonic and radiative decays. However in doing so we better
keep the advice of Maitre Le Yaouanc at heart
\cite{LEYAOUANC}: the quark model
is a wonderful device to develop our intuition and check our
conjectures; yet at the same time its proper application
involves an artful evaluation of past experiences rather than a
rigid scientific procedure -- exactly like it is with
good French cooking!

\section{\bf Status Quo}

One can discuss the status of {\em hadronic beauty production}
like the well-worn story about the optimist/pessimist looking
at a half-full/half-empty glass of wine: while CDF measures
\cite{CDF}
a larger beauty cross section than predicted, I am not yet
convinced that this represents an alarming problem for theory;
for proper perspective one should also keep in mind that the
observed $b-\bar b$ correlations seem to follow the
expected pattern.

The {\em `average' beauty lifetime} has been measured to be
close to 1.5 psec establishing that indeed $|V(cb)| \sim
{\cal O}(\sin ^2\theta _C)$ rather than
$\sim {\cal O}(\sin \theta _C)$. This is a very important
observation with far-reaching implications as mentioned above.
Yet it should now assume the place of honour it richly
deserves in the `museum of important measurements'.
For once you have
established that the `typical' beauty lifetime is indeed
around 1 psec, any numerically more precise measurement
of the average lifetime in an a priori unknown cocktail
of beauty hadrons can certainly advance bragging rights,
but not our understanding of the physics involved. For that
purpose one has to extract the {\em lifetimes of specified
beauty hadrons}, i.e. charged vs. neutral mesons vs.
baryons. In the table below I juxtapose theoretical predictions
obtained within QCD through an expansion in $1/m_b$ with
present data \cite{STONE}:

\begin{center}
\begin{tabular} {|l|l|l|}
\hline
& &\\
&QCD ($1/m_b$ expansion) &data \\
& & \\
\hline
&  & \\
$\tau (B^-)/\tau (B_d)$ & $1+
0.05(f_B/200\, \MeV )^2
[1\pm {\cal O}(10\%)]$ &$1.12 \pm 0.09$ \\
& & \\
\hline
& & \\
$\bar \tau (B_s)/\tau (B_d)$ &$1\pm {\cal O}(0.01)$
&  $ \sim 1$ \\
& & \\
\hline
& & \\
$\tau (\Lambda _b)/\tau (B_d)$&$\sim 0.9$ & $0.75\pm 0.12$\\
& & \\
\hline
\end{tabular}
\end{center}
\vspace{0.5cm}
A few comments are in order here:

\noindent
$\bullet$ To (formal) order $1/m_b^3$ one predicts
unequivocally $\tau (B^-) > \tau (B_d)$. The quoted uncertainty
reflects corrections of order $1/m_b^4$ that have not been
determined; also their sign is uncertain.

\noindent
$\bullet$ It is very hard to see theoretically
how $\tau (\Lambda _b)$ could differ from $\tau (B_d)$ by more
than 10 \%.

\noindent
$\bullet$ No theoretical desaster has occurred; i.e., while
the data are not sufficiently precise to establish the
predictions, they are quite consistent with them. This
constitutes a non-trivial semi-quantitative success,
since the lifetime differences in the beauty sector indeed
are much smaller than in the charm sector.

\noindent
$\bullet$ {\em If} future data establish a significant deviation
from the predictions we would be faced with a serious theoretical
conundrum. There is no `model' parameter that could freely be
adjusted. One would quite naturally suspect that the deficiencies
of our expansions arise mainly in the description of non-leptonic
decays. There are actually two variants for such a conjecture:
(a) $1/m_Q$ expansions are of no use in  non-leptonic decays in
general although they can be employed with profit in semileptonic
decays. (b) Alternatively -- and a detailed analysis of the concept
of quark-hadron duality seems to point in that direction -- one would
conclude that the $1/m_b$ expansion numerically converges more
slowly in non-leptonic $b\ra c$ than in semileptonic $b\ra c$ and in
all $b\ra u$ transitions. (One should note that the energy release
is larger in $b\ra u$ than in $b\ra c$ transitions.)

Time-dependant manifestations of $B^0-\bar B^0$ {\em oscillations}
have been analysed at LEP in a most impressive way:
$$\Delta m(B_d)=0.509 \pm 0.046 \; (psec)^{-1}\eqno(1)$$
$$ x_s\equiv \frac {\Delta m}{\Gamma}|_{B_s} > 2.5 \eqno(2)$$
Once the top quark mass is known with good accuracy one can
infer some information on $|V(td)|$ from a measurement of
$\Delta m(B_d)$. Yet keeping in mind that
$\Delta m(B_d)$ depends roughly linearly on $m_{top}$ in the
relevant range and that hadronic uncertainties arise -- as
expressed through the quantity $B_Bf_B^2$ --, it seems to me
that $\Delta m(B_d)$ is not quite, but almost as well
measured as required. Based on the
SM bound $x_s > 5$ \cite{ALI}
one concludes that improvements in the
LEP measurements anticipated for the next two years
have the potential to provide strong evidence for the
intervention of NP beyond SM. In the context of
$B_s-\bar B_s$ oscillations one should also point out
that there are actually two lifetimes,
namely one for a $B_{s, short}$ and one for
a $B_{s, long}$ with their difference estimated to be
\cite{GAMMABS}
$$\frac{\tau (B_{s,long})-\tau (B_{s,short})}
{\bar \tau (B_s)}\simeq 0.18 (\frac{f_{B_s}}{200\, \MeV})^2
\eqno(3)$$
It is intriguing to note that this delicate phenomenon is
expected to yield the largest lifetime difference among $B$
mesons. This could remain an academic observation since a
difference as predicted by eq.(3) is presumably too small
to be observable. On the other hand the calculation underlying
this estimate for $\Delta \Gamma (B_s)$ is not `gold-plated'
and could conceivably underestimate it significantly. Therefore
one should endeavour to search for separate
$B_{s,long}$ and $B_{s,short}$ lifetimes even if one can
attain sensitivity only for a 50\% or 100\% difference.
The cleanest method for such an analysis is probably
to determine the $B_s$ lifetime in two different classes
of decay channels: e.g.,
$$\tau (B_s\ra l \nu D_s^{(*)})-
\tau (B_s\ra \psi \phi )\simeq
\frac{1}{2}[\tau (B_{s,long})-\tau (B_{s,short})]\eqno(4)$$

{\em The KM parameters} $|V(cb)|$ and $|V(ub)|$ can best
be determined in semileptonic decays. There are two
rather reliable methods available:

\noindent (a) One can
compare the measured inclusive
semileptonc width for B mesons with the theoretical
expression through order $1/m_b^2$; that way one
deduces \cite{SUV}
$$|V(cb)|_{incl} \simeq 0.0415 \pm 0.002 \pm experimental\;
errors \eqno(5)$$
where I have specified the present {\em theoretical}
uncertainty only. It is so small, since
$\Gamma _{SL}(B)$ turns out to
depend mainly on the mass difference
$m_b-m_c$, rather than on $m_b$ separately; the heavy quark
expansion allows to determine this difference quite
precisely from the observed values of the $D^{(*)}$
and $B^{(*)}$ masses.

\noindent (b) Measuring the exclusive
channel $B\ra l \nu D^*$ as a function of the
energy transfer and extrapolating to the zero recoil
point, one extracts $F_{B\ra D^*}(0)|V(cb)|$;
the most recent CLEO data yield \cite{CLEO}:
$F_{B\ra D^*}(0)|V(cb)| = 0.0351\pm 0.0019 \pm
0.0018 \pm 0.0008$. From Heavy Flavour Symmetry one infers:
$F_{B\ra D^*}(0)=1+{\cal O}(\alpha _S/\pi)+
{\cal O}(1/m^2_{b,c})$. It had been claimed
\cite{NEUBERT} that the
non-perturbative corrections reduce this formfactor by
merely 2-3 \%. {\em If} true it would be very surprising:
for the $1/m_Q^2$ terms should be dominated by the
charm mass leading to corrections of order
$(\mu _{had}/m_c)^2\sim 10 \%$. Very recently it has been
shown through an application of $1/m_Q$ expansions in the
small velocity limit that the pre-asymptotic corrections
are actually
larger than previously claimed and in agreement with the
simple expectation sketched above \cite{SUV1}:
$F_{B\ra D^*}(0)\simeq 0.89 \pm 0.03$ and therefore
$$|V(cb)|_{excl}\simeq 0.0394 \pm 0.0034 \eqno(6)$$
in good agreement with the value obtained from the
inclusive analysis, eq. (5).

The situation is much more uncertain concerning
$|V(ub)/V(cb)|$: even once the exclusive mode
$B\ra l\nu \rho$ has been measured, it will pose a
formidable challenge to theory to extract a value
for $|V(ub)/V(cb)|$ as emphasized by
Le Yaouanc \cite{LEYAOUANC}. As
far as {\em inclusive} semileptonic $B$ decays are
concerned, we are in a better position -- also
because clear evidence for $b\ra u$ transitions has
been found there. In analysing the inclusive lepton
spectrum there is wide-spread consensus on the
procedure to be followed: one picks one's favourite
model -- $AC^2M^2$, $ISGW$ etc. --, fixes its shape
parameters from fitting the spectrum in
$B\ra l\nu X_c$ and applies it to the endpoint region
from where one extracts $|V(ub)/V(cb)|$. Considerable
progress has been made recently in putting this
procedure onto a firmer theoretical foundation: it
has been noted that a properly re-defined
$AC^2M^2$ model represents a faithful (though not
unique) realization of QCD for $b\ra u$ (but not
for $b\ra c$) transitions \cite{ROMAN}; furthermore it has been
noted that the {\em shape} of the lepton spectrum in
$B\ra l\nu X_u$ can be deduced from the photon
spectrum in radiative $B\ra \gamma X_s$
decays \cite{MOTION}. Rather
reliable extractions of $|V(ub)/V(cb)|$ will thus become
possible in the future. Yet for the moment considerable
uncertainties exist. I will use
$$|V(ub)/V(cb)| \sim 0.05 - 0.11\; ; \eqno(7)$$
however I am {\em not} convinced that even this range properly
reflects the uncertainty.

The quantity $\sin 2\beta$ with $\beta$ denoting one of the
angles in the \KM triangle determines the SM CP asymmetry
in the decay $B_d/\bar B_d \ra \psi K_S$. Its size can be
inferred from the measured value for the ratio
$\epsilon _K/\Delta m(B_d)$ with fairly little sensitivity
to the value of $m_t$ -- albeit with considerable
theoretical uncertainties concerning the size of
$\Delta B=2$ (and $\Delta S=2$) matrix elements:
$$\sin 2\beta \simeq 0.33\times UNC \eqno(8a)$$
$$UNC\simeq \left( \frac{0.045}{|V(cb)|}\right) ^2
\left(\frac{0.72}{x_d}\right)
\left(\frac{\eta ^{(B)}_{QCD}}{0.55} \right)
\left(\frac{0.62}{\eta ^{(K)}_{QCD}}\right)
\left( \frac{2B_B}{3B_K}\right)
\left(\frac{f_B}{160\, \MeV}\right) ^2\eqno(8b)$$
with $x_d\equiv \Delta m(B_d)/\Gamma _B$.
Our present information on the shape of the \KM
triangle is summarized in Fig.~\ref{F4A}:
the top of the triangle
has to lie in the shaded area. One reads off without
ado that none of the three angles is particularly
small; the corresponding CP asymmetries are then
predicted to fall in the range of few
$\times$ 10\%!
\begin{figure}
\begin{center}
\mbox{\epsfig{file=7207fg4.eps,height=6cm}}
\end{center}
\caption[]{Shape of the KM triangle inferred from present
phenomenology}
\label{F4A}
\end{figure}

\section{\bf The Next Five Years}

Several actors will carry the drama of beauty physics
during that period, namely CLEO II, the LEP
experiments, CDF and maybe D0.

At the FNAL collider hadronic b production in the
central region wil be studied more precisely and
conclusively.

Lifetime measurements will be refined and extended
where the following levels of accuracy might be attained:

$$\delta [\tau (B^-)/\tau (B_d)]=\pm (3-5)\%, \;
\delta [\tau (B_s)/\tau (B_d)]=\pm (5-7)\%, $$
$$\delta [\tau (\Lambda _b)/\tau (B_d)]=\pm (5-7)\%, \;
\delta [\tau (\Xi _b)/\tau (\Lambda_b)]=\pm (20)\%, \;
\delta [\tau (\Xi ^0_b)/\tau (\Xi ^-_b)]=\pm (20)\% .
\eqno(9)$$
While the precision of these measurements presumably
will not suffice to firmly establish the lifetime pattern
as predicted in QCD, they will at the very least be
valuable in that they probe for potential trouble.

Let me make a side remark on a `cute' system, namely
$B_c\equiv (b\bar c)$. It provides a rich spectrum of
excitations that are calculable; the Isgur-Wise
function for the decay $B_c\ra l \nu \psi$ can be
computed as well; its overall decays reflect an
intriguing interplay of three classes of transitions,
namely the decay of the $b$ quark, that of the
$\bar c$ (anti)quark and the `Weak Annihilation'
of $b$ and $\bar c$. Ignoring the latter one would
naively expect the $\bar c$ decay to dominate:
$\Gamma (B_c\ra B+X_s) \sim \Gamma (D^0)\simeq
(4\cdot 10^{-13}\, sec)^{-1} >
\Gamma (B_c\ra X_{c\bar c})\sim \Gamma (B)\simeq
(1.5\cdot 10^{-12}\, sec)^{-1}$. It has been suggested
\cite{QUIGG}
that for a tightly bound system like $B_c$ the
decay width should be expressed not in terms of the
usual quark masses $m_b$ and $m_c$, but instead in terms of
effective quark masses reduced by the binding energy
$\mu _{BE}$. {\em If so}, then the $B_c$ decay width
would be reduced significantly since due to
$\Gamma \propto m_Q^5$ one would find
$\Delta \Gamma /\Gamma \sim 5\mu _{BE}/m_{b,c}$;
even more significantly, beauty decays would become
more frequent among $B_c$ decays than charm decays since
the binding energy constitutes a higher fraction of
$m_c$ than of $m_b$. However such a conjecture is
fallacious! For it has been shown \cite{BUV}
that the leading non-perturbative corrections
arise at the $1/m_{b,c}^2 \leq 10\, \%$ level; the naive
estimate is thus basically correct, i.e.
$\tau (B_c) \sim 3\cdot 10^{-13}\, sec$ with charm
decays dominating over beauty decays.

$B_s-\bar B_s$ oscillations will be probed with a
sensitivity for $x_s$ up to $\sim$ 5--10. A
positive signal, at least in the lower range, would signal
the intervention of NP. Furthermore a $\sim$ 30 \%
difference in $\tau (B_s\ra \psi \phi)$ vs.
$\tau (B_s\ra l \nu D_s^{(*)})$ can be searched for. If
found, it would presumably provide us with a valuable
lesson on QCD rather than on NP.

The SM parameters will be determined with improved
accuracy: $\delta m_t\simeq 10\, \GeV$,
$\delta |V(cb)|< 5\, \%$,
$\delta |V(ub)|\simeq 10-20\, \%$
and $\delta |V(td)|< 30\, \%$ seem to be
achievable. It is assumed here that $|V(td)|$
has been extracted from a measurement of
$BR(K^+\ra \pi ^+ \nu \bar \nu)$.

Measurements or bounds on $B\ra \pi \pi$ vs.
$B\ra K \pi$ vs. $B\ra K \bar K$ will yield some clues
on the weight of final state interactions, Penguins
etc.

The hadronic parameters $f_D$ and $f_{D_s}$ might be
extracted from $D^+, D_s \ra \mu ^+\nu$ with `useful'
errors. This would allow us to calibrate the findings of
QCD simulations on the lattice and to extrapolate to
$f_B$ in a controlled way.

In Fig.~\ref{F4B} I show the KM triangle resulting
from $m_t=160\pm 10\, \GeV$, $|V(ub)|/|V(cb)| =
0.08\pm 0.01$ and $f_B=140\div 240\, \MeV$. There
emerge two disjoint shaded areas now, i.e. there are
two disjoint allowed regions in parameter space --
unless one can decide between $f_B\leq 170\, \MeV$
and $f_B\geq 210\, \MeV$ (for the assumed values of the
SM parameters). Those two regions lead to quite different
predictions on $x_s$, namely $x_s\simeq 8-10$ for the
left shaded region and $x_s\simeq 17- 25$ for the
right one (where I have assumed
$(f_{B_s}/f_{B_d})^2 = 1.1$).
\begin{figure}
\begin{center}
\mbox{\epsfig{file=7207fg5.eps,height=6cm}}
\vspace*{-1cm}
\end{center}
\caption[]{Shape of the KM triangle after future measurements
of V(ub)/V(cb) and $m_t$}
\label{F4B}
\end{figure}

$B\ra K^*l^+l^-,\, K l^+l^-$ might be observed,
$B\ra \gamma +${\em higher K resonances} will be studied
and the inclusive reaction $B\ra \gamma X_s$ measured, the
last item by CLEO II; they will search also for direct
CP violation in $B_d\ra K^+\pi ^-$ vs.
$\bar B_d\ra K^-\pi ^+$ etc.

\section{\bf The Heroic Period}

Great heroic deeds -- in particular in uncovering CP
violation -- are waiting to be achieved by the bold one,
and there is a considerable list of potential heroes:
HERA-B, the SLAC and KEK asymmetric B factories,
CDF/D0 in the Main Injector era and CLEO III. Of course
we all understand that when you see a list of potential
heroes, you are at the same time looking at
scapegoat candidates. Picture yourself in a big
European football match: regulation time is almost
over, the score is tied and you get a penalty kick.
While you trot over to the penalty box to take the kick
you realize that there are only two
extreme outcomes for your undertaking: you score,
you are the hero and get carried off the field on the
shoulders of your peers -- or you miss, you are the
scapegoat and will always be remembered by your failure,
irrespective of your achievements before or after. It seems
to me that HERA-B and the two asymmetric B factories are
in that precarious situation \footnote{Fortunately there are
better than even odds that you score.}.

Yet before addressing the
quest for heroism, let me evaluate less romantic
endeavours.

Hadronic beauty production in the forward region might be
studied with one eye on the topic of
diffraction \cite{EGGERT} and the
other on a search for asymmetries in $B$ vs. $\bar B$
{\em production}.

Lifetimes will be determined with high accuracy, say of order
1 \% or so. This will allow to establish and measure the
lifetime differences as predicted in QCD.

$B_s-\bar B_s$ oscillations will be probed further with the
sensitivity in $x_s$ reaching values of
$\sim$ 15 and maybe even 20. While {\em present} SM
phenomenology admits $x_s$ in this whole range above
$x_s > 5$, this will have changed by that time! In the
hypothetical, yet generic scenario exhibited in
Fig.~\ref{F4B}, $x_s$ values in the window between
$\sim$ 10 and $\sim$ 17 would require NP.

The SM parameters might get measured with an accuracy
approaching what can realistically ever be achieved,
namely $\delta m_t\simeq 3\, \GeV$,
$\delta |V(cb)|= 1\, \%$,
$\delta |V(ub)|\simeq few\, \%$
and $\delta |V(td)|= 10\, \%$. In order to
attain such levels of precision one will have to
measure the photon spectrum in
$B \ra \gamma +X_s$ and analyse inclusive as well as
exclusive semileptonic decays {\em separately}
for $B^-$ and $B_d$ mesons (for $|V(ub)|$), or
measure $BR(K^+\ra \pi ^+\nu \bar \nu)$ and/or
$\Delta m(B_s)$ vs. $\Delta m(B_d)$ with
$\sim$ 20 \% accuracy (for $|V(td)|$).

With some luck, $f_B$ might get determined through
$B\ra \tau \nu$ decays. Even without that gift the
KM triangle might already be overconstrained {\em before}
meaningful CP studies are undertaken in $B$ decays. Without
loss of generality one can normalize the baseline to unity.
Together with $|V(ub)|/|V(cb)|$ and
$|V(td)|/|V(cb)|$ the lengths of the three sides are
known thus defining the triangle from which one reads off the
value of the angle $\beta$. Employing eqs.(8) one can
then determine the value for $B_Bf_B^2$ required to reproduce
$\epsilon _K/\Delta m(B_d)$. Inserting the
value thus inferred for $B_Bf_B^2$ together with that
for $|V(td)|$ and $m_t$ into the SM expression for
$\Delta m(B_d)$, one can finally check whether the observed
magnitude for $\Delta m(B_d)$ is reproduced.

Now to the hoped for heroic deeds.
The CP asymmetry in $B_d\ra \psi K_S$ yields (within SM)
the quantity $\sin 2\beta$ unequivocally. If $|V(td)|$
and/or $f_B$ are known {\em independantly}, then a difference
in the value of $\sin 2\beta$ as {\em measured} in
$B_d/\bar B_d \ra \psi K_S$ and as {\em inferred}
from the KM triangle establishes the intervention of
NP. Otherwise the normalized triangle is defined through
its baseline of unit length, the side
$|V(ub)/\lambda V(cb)|$ and the angle $\beta$, i.e.
purely through measurements in the $B$ system.

{}From the observed CP asymmetry in
$B_d/\bar B_d\ra \pi ^+\pi ^-$ one deduces the angle
$\alpha$ \footnote{Final state interactions in general and
`Penguin' effects in particular could modify the relation
between $\alpha$ and the CP asymmetry; in that case
$B_d\ra \pi ^+\pi ^-$ might well exhibit also direct
CP violation that could in principle be observed at CESR.}.
If this measured value of $\alpha$ differs significantly from
the value that is inferred from the KM triangle, NP has
finally been found in an unequivocal way.

In principle one can measure
the third angle $\gamma$ in two different
ways, namely

\noindent
$\bullet$ through a difference in $\Gamma (B^+\ra D^0/\bar D^0K^+)$
vs. $\Gamma (B^-\ra D^0/\bar D^0K^-)$; it represents direct CP
violation and is proportional to $\sin \gamma$
\cite{PAIS,GRONAU};

\noindent
$\bullet$ through a difference in $B_s\ra K_S\rho ^0$ vs.
$\bar B_s\ra K_S\rho ^0$; it involves $B_s-\bar B_s$ oscillations
and depends on $\sin 2\gamma$ \footnote{It is at least amusing
to note that this asymmetry would vanish for $\gamma = 90^{\circ}$
whereas the previous one would be maximal.}.

While the first of these two methods definitely
has some promise and relevance since
it can be undertaken at a symmetric $B$ factory,
the second one has
not too much of it,
I believe. The CP asymmetries in $B_d\ra \psi K_S$ and
$B_d\ra \pi ^+\pi ^-$ will be measured sooner and with greater
accuracy than in $B_s\ra K_S\rho ^0$, and $\gamma$ will be known
as $\gamma = 180^{\circ} -\alpha -\beta$ within SM! The real
motivation in determining $\gamma$ independantly is to search
for a manifestation of NP. Yet this is a more promising undertaking
in a mode that commands a higher branching ratio, a more striking
signature and a cleaner interpretation: the modes
$B_s\ra \psi \phi$ and $B_s\ra \psi \eta$ fit the
bill \cite{BS}, in
particular since SM predicts a very small CP asymmetry for it
($\leq 2\%$). Comparing $B_s\ra \psi \phi$ with
$\bar B_s\ra \psi \phi$ thus represents a practically zero
background search for NP!

As I will stress later on, it will be mandatory to measure
the angles $\alpha$ and $\beta$ as precisely as possible --
for the ultimate analysis. Yet at intermediate stages one
can perform a very valuable and promising research program
even if one has to restrict oneself to the HERA-B `menu',
namely a study of three reactions:
(i) $B_d/\bar B_d \ra \psi K_S$, (ii) $B_s-\bar B_s$
oscillations and
(iii) $B_s/\bar B_s\ra \psi \phi , \, \psi \eta$.
To sketch what can be achieved through such a
menu I assume that $|V(cb)|$, $|V(ub)|$ amd $m_t$ have
been determined with good accuracy. {\em If} $|V(td)|$
has been extracted from $BR(K^+\ra \pi ^+ \nu \bar \nu)$
by that time, the KM triangle will have been defined. One reads
off the value predicted for the angle $\beta$ and using eqs.(8)
one infers the required size of $B_Bf_B^2$. Then one is in a
position to perform four sensitive tests of SM:
$(\alpha )$ Using $m_t$ and the inferred value of $B_Bf_B^2$,
one can check whether the observed value of
$\Delta m(B_d)$ is reproduced.
$(\beta )$ Compare $\sin 2\beta$ as deduced from the KM
triangle with the CP asymmetry actually observed in
$B_d\ra \psi K_S$. $(\gamma )$ Search for $B_s-\bar B_s$
oscillations. As mentioned in the preceding section,
at that time $x_s$ will be predicted to fall into
one or two
relatively narrow windows.
$(\delta )$ Any observation of a CP
asymmetry in $B_s\ra \psi \phi$ establishes the intervention
of NP.

\noindent {\em If} on the other hand $BR(K^+\ra \pi ^+ \nu \bar \nu )$
has not been observed (yet) with sufficient reliability,
then one has to use the observed CP asymmetry in
$B_d \ra \psi K_S$ as the final defining element in the
KM triangle; three tests then remain:
$(\alpha )$ Is $\Delta m(B_d)$ reproduced with the value for
$B_Bf_B^2$ as deduced from $\beta$?
$(\beta )$ Probe $B_s-\bar B_s$
oscillations.
$(\gamma )$ Search for a CP
asymmetry in $B_s\ra \psi \phi$.

Finally I would like to comment on how {\em production}
asymmetries, if present, can be employed with great profit.
As we have heard from Berger, differences in the production of
$B$ and $\bar B$ mesons in $p \bar p$ collisions are expected to
occur at a level of maybe up to 10 \%, but only in the extremely
forward (or backward) region. Surprises, of course, could happen,
i.e. larger production asymmetries could occur in wider parts of
phase space, and I would rate them as mere sins against orthodoxy
rather than outright heresies. As such one should -- in the
spirit of `peccate fortiter' -- make use of them once they occur,
irrespective of what they mean for the pure doctrine; for they
would remove the need for an independant flavour tag. Consider a
beam of $N+\bar N$ neutral $B$ mesons; its decay into a
{\em CP eigenstate} like $\psi K_S$ as a function of proper
time $t$ is given by
$$rate\, (B^{neut}\ra \psi K_S;t)\propto e^{-\Gamma t}
(1+\frac{N-\bar N}{N+\bar N}\cdot \sin 2\beta \cdot
\sin \Delta mt)\eqno(10)$$
If there is a production asymmetry -- $N\neq \bar N$ --
then CP violation becomes observable {\em without} an
independant flavour
tag. Furthermore the quantity $(N-\bar N)/(N+\bar N)$ can be
tracked independantly through decays into
{\em CP non-eigenstates}:
$$rate\, (B^{neut}\ra \psi K^-\pi ^+;t)\propto
e^{-\Gamma t}(1+\frac{N-\bar N}{N+\bar N}\cdot
\cos \Delta mt)\eqno(11a)$$
$$rate\, (B^{neut}\ra \psi K^+\pi ^-;t)\propto
e^{-\Gamma t}(1-\frac{N-\bar N}{N+\bar N}\cdot
\cos \Delta mt)\eqno(11b)$$
That way one can check also for a detector bias.

\section{\bf LHC -- The Ultimate Measurements}

I am going to argue now that the LHC will satisfy
essential needs in a complete program of beauty physics --
even if all prior enterprises have succeeded as well as
promised by their proponents. This judgement is based on
three reasons, of which the last one is truly major:

(1) $x_s > 20$ is quite possible (below I will cite
some relevant examples); this range seems to be well beyond the
reach of any pre-LHC enterprise.

(2) Detailed measurements of
$b\ra s\, l^+l^-,\, d\, l^+l^-$
transitions will require statistics that
can be accumulated only at the LHC.

(3) As far as CP violation is concerned there are three
possible outcomes for a dedicated program:

\noindent {\bf (A):} {\em No CP violation is observed in $B$ decays!}
To be more specific:
it is found that CP asymmetries in relevant $B$
decays are at most a few percent.
Since there exist three families, $B^0-\bar B^0$ oscillations
proceed speedily and based on what we already know of the
KM parameters, I regard this scenario as the least likely
outcome. Yet in that case we would have established that the
KM mechanism is {\em not} behind
$K_L\ra \pi \pi$, that it is not even a significant factor there!
It would force us to abandon the SM paradigm of CP violation
and to attribute the {\em observed} CP violation in $K_L$ decays
to the intervention of as yet unidentified NP, with two
immediate consequences: (i) It would lead to the formulation
of a baffling theoretical puzzle, namely `Why is there no
significant KM source for CP violation?'. (ii) On the
experimental side it would lend topical urgency to
searches for different manifestations of CP violation in
light-quark systems, such as the electric dipole moments of
neutrons or electrons, the transverse muon polarization
in $K_{\mu _3}$ decays, $BR(K_L\ra \pi ^0l^+l^-)$ etc.

\noindent {\bf (B):} {\em Large CP asymmetries are found --
but they are not (quite) consistent with the KM
predictions!} This scenario -- which I regard as the most
likely outcome -- will represent indirect, but unequivocal
evidence for the intervention of NP in $B$ transitions.
Taken together with measurements of $m_{top}$,
$\Delta m(B_s)$, $BR(b\ra s \gamma)$, $BR(b\ra sl^+l^-)$
and hopefully $BR(K^+\ra \pi ^+ \nu \bar \nu)$ and
$BR(b\ra d\gamma )$ it would provide us even with some definite
clues about the nature of the NP involved. It would also allow
us to check to which degree the KM mechanism contributes to
$K_L\ra \pi \pi$.

\noindent {\bf (C):}
{\em Large CP asymmetries are found without evidence for
NP -- even when extrapolating down to $K_L$ decays!}
Also such a scenario would provide us with the seeds for more
profound knowledge, since it addresses one of the most central
mysteries of the Standard Model, namely the generation of
fermion masses. After a comprehensive analysis of $B$ decays
the KM matrix will be known. Keep in mind that the KM
matrix which arises due to a non-trivial diagonalization
of the up-type and the down-type quark matrices contains
unique information $over$ and $above$ the six quark mass
eigenvalues. Some attempts have been undertaken to construct
the KM matrix in terms of the six quark masses by
conjecturing a simple form of the quark mass matrices
at a high (SUSY) GUT scale and then evolve it down to the
hadronic scales probed in heavy flavour decays.
Comprehensive studies of this kind have recently been given in
refs.\cite{RRR,RABY};
I summarize their results for $m_{top}=180\, \GeV$
in a way that is convenient for my discussion
\footnote{It takes, of course, the trained eye of a theorist
to discern the simple pattern underlying these values for the
KM parameters.}:

\begin{center}
\begin{tabular} {|l|l|l|l|l|l|l|l|l|}
\hline
&A &B &C &D &E &I &II &III \\
\hline
$|V(ub)/V(cb)|$ & 0.06 & 0.062 & 0.068 & 0.059
&0.089 &0.046 &0.059 &0.071 \\
\hline
$x_s/x_d$ & 25 & 25 & 22 & 26 & 35 &  &  &  \\
\hline
$\sin 2\beta$ & 0.52 & 0.54 & 0.58 & 0.51 & 0.71
&0.39  &0.49  &0.59  \\
\hline
$\sin 2\alpha$ & -0.15 & -0.11 & 0.30 & -0.40 & -0.62
&-0.32 &-0.46 &-0.14 \\
\hline
$\sin 2\gamma$ & 0.65 & 0.71 & 0.30 & 0.81 & 0.99
&0.66 &0.84 &0.70 \\
\hline
$\sin \gamma$ & 0.94 & 0.94 & 0.99 & 0.89 & 0.75
&0.93 &0.88 &0.93 \\
\hline
\end{tabular}
\end{center}
where I have assumed $[B_sf^2(B_s)]/[B_df^2(B_d)]=1.1$
to translate $|V(ts)/V(td)|^2$ into $x_s/x_d$. The symbols
$A-E$ and $I-III$ refer to different classes of mass matrices
analysed in ref.\cite{RRR} and
\cite{RABY}.  The details are not important here,
and I anticipate considerable theoretical evolution to take place
over the next few years; but I want to use these numbers to
illustrate important benchmarks for the ultimate measurements:

\noindent $\bullet$ One better be able to probe $B_s-\bar B_s$
oscillations up to $x_s\sim 40$ or even beyond.

\noindent $\bullet$ Both $\sin 2\beta$ {\em and}
$\sin 2\alpha$ have to be measured -- that is non-negotiable!

\noindent $\bullet$ While one aims for a 20-30 \% accuracy for the
first and second round measurements, the goal has to be to finally
achieve a 5\% precision or better to exploit the discovery
guarantee to the fullest and to distinguish between the different
scenarios.

\noindent $\bullet$ In addition one has to strive to perform
as precise measurements as possible for $B\ra \gamma \rho, \,
\gamma \omega$ vs. $B\ra \gamma K^*$,
$B\ra l^+l^- \pi / \rho / \omega$ vs.
$B\ra l^+l^- K/K^*$ (including the $l^+$ vs. $l^-$ spectra),
$B\ra \mu ^+ \mu ^-, \, (\tau ^+ \tau ^-)$ etc.

It is obvious that only the LHC has the `statistical' muscle to
accumulate the necessary sample sizes. The challenge is: can the
LHC develop the required `systematic' brain!

\section{\bf Outlook}

A long and arduous, if not outright tortuous journey lies ahead
of us. What should keep us going is the realization that the insights
that will be gained from a comprehensive and detailed program of
beauty physics

\noindent $\bullet$ are of fundamental importance,

\noindent $\bullet$ cannot be obtained any other way, and

\noindent $\bullet$ cannot become obsolete.

Finally I can explain why Mont St. Michel represents such
a fitting allegory for the research program we have been
discussing here. Some buildings, like the church or
`La Merveille', are more spectacular than others, but it
is the whole ensemble that makes Le Mont --  like beauty
physics -- so unique and irresistible, and it is only because
of the whole (data) base and foundation
that the spire crowning the rock
can point -- to the New Standard Model!

\vspace*{0.5cm}

{\bf Acknowledgements:} \hspace{.4cm}
This work was supported by the National Science Foundation under
grant number PHY 92-13313. I thoroughly enjoyed the lively
and stimulating atmosphere created by the organizers of
the meeting, P. Schlein and Y. Lemoigne, and I learned
a great deal from Lemoigne's comments on art and history.
There were many conversations from which I benefitted, in
particular with E. Berger, A. Le Yaouanc and T. Nakada.

\vspace*{3cm}
\end{document}